\documentstyle[twocolumn,aps,prd]{revtex}

\begin{document}

\title{Summary Talk on Cosmology and  Gravitation \\
XXII Brazilian National Meeting on Particles and Fields}

\author{M.J. Rebou\c{c}as}

\address{Centro Brasileiro de Pesquisas F\'{\i}sicas\\
Departamento de Relatividade e Part\'{\i}culas\\
Rua Dr Xavier Sigaud, 150\\
22290-180 Rio de Janeiro -- RJ, Brazil \\
E-mail: reboucas@cbpf.br} 

\maketitle

\begin{abstract}

The plenary lectures, parallel talks, oral presentations and panel 
contributions on Cosmology and Gravitation presented during the
XXII Brazilian National Meeting on Particles and Fields are briefly 
reviewed. Some remarks on the area are also presented.

\end{abstract}

\section{Overview and Introduction}

After four days of listening to two plenary lectures, two parallel 
talks, a general public lecture, seventeen oral presentations, and 
reading more than fifty panels related to Cosmology and Gravitation, 
my chief impression is that the XXII Brazilian National Meeting on 
Particles and Fields (XXII BNMPF) was an noticeable scientific 
meeting, which covered a great variety of issues on Gravitation
and Cosmology. It brought together perhaps most of the researchers 
and graduate students that have worked in this area in Brazilian
Universities and Research Institutes. The sessions were very
well attended with a fair number of clarifying questions. In
general the speakers explained the underlying ground, approaches
and results in a clear manner and kept to the time limits. I
thank to them all for their quite good talks.

{}From this and the previous meetings I support the view that 
if one considers the contributions presented in, e.g., two 
successive BNMPF meetings they are rather  representative of 
the researches that have been done in this area in Brazil, and
most of the contributions reveal a clear emphasis on theoretical 
and (or) mathematical aspects of Gravitation and 
Cosmology~\cite{IWaga1}. 

The researches in modern cosmology can be broadly grouped 
in five domains:
\begin{enumerate}
\item[i.]
Astrophysical cosmology; 
\item[ii.] Observational cosmology; 
\item[iii.] 
The early universe;  
\item[iv.] 
Theoretical and mathematical cosmology;
\item[v.] Cosmic topology (more recently).  
\end{enumerate}
Most of the contributions to the XXII BNMPF (organized by 
the Brazilian Physical Society - SBF) were in the last three 
wide families. On the other hand, as far as I am aware a 
great number of the observationally-oriented works on cosmology 
are often presented in the annual meeting of the Brazilian 
Astronomical Society (SAB). 

I do not share the view that theoretical (and mathematical) 
cosmology should be strictly related to the observed universe, 
otherwise it becomes a pointless exercise. In my evaluation 
the frontier between theoretical (and mathematical) cosmology 
and more observationally-oriented cosmology is not and 
should not be well-defined, and it is very important that 
we have people working on both general aspects of 
modern cosmology. However, I understand that a closer 
interaction between theoretical and 
observationally-oriented cosmologists is more than desirable
--- in my view it is recommendable --- for the benefit of  
both (physicist and astophysicist) communities. The recent 
workshop called {\em New Physics in the Space\/} is certainly 
a good example in this direction~\cite{Opher1}.
   
A considerable number of abstracts were submitted to the
organizers. I envisage how hard was to 
select between oral and panel presentations. A number 
of relevant subjects seemed to them to be of interest 
to motivate oral presentations, but no doubt 
other organizers would have made a quite 
different, and perhaps equally good, selection. 
I will summarize the presented contributions below, 
with apologies to their authors for any unintentional 
misapprehension and (or) misrepresentation.

\section{Plenary and Parallel Talks}

The General Relativity (GR) predicts the existence of 
gravitational waves, and analyses of experimental data obtained 
{}from observations of binary star systems seem to give a clear
evidence that these systems loose energy through gravitational
radiation~\cite{TaylorRev}.
 
Gravitational waves are predicted not only by GR but also
by other metric gravitational theories of some generality and 
scope. However, the characteristics of gravitational waves
such as propagation speed, multipolar structures and polarization 
states vary from one theory to another. Thus, it is expected
that its detection will provide a rather relevant tool to 
select the most suitable theory of gravitation~\cite{Will}.
Further, the direct observation of gravitational waves
also offer the possibility of testing GR (as well as other
metric theories) in a strong field limit, where the effects
are not merely a correction of Newtonian Gravitation.
Obviously at Earth the waves are expected
to be weak perturbations, however they can in principle 
provide information on the conditions at their strong field 
sources~\cite{Barish1}.  
Further, the technology needed to detect as well as the 
analysis of gravitational waves is expected to open a new window 
for observation of the universe, thus giving rise to a remarkable
period in astronomy and cosmology.

Barry C. Barish (California Institute of Technology) delivered 
a clear plenary lecture in the XXII BNMPF, in which he 
presented a good review of the new generation of detectors with 
suitable sensitivities for detecting gravitational waves from a 
variety of catastrophic events, such as the gravitational collapse 
of stars or coalescence of compact  binary systems. He pointed out 
that more than forty years after the beginning of the search for 
gravitational waves, several resonant-mass wave detectors are 
monitoring the most strong potential sources of such waves 
(in the $\approx k\,Hz$ band) in our galaxy and in our local 
group. The gravitational wave detection of low frequencies 
(in the $\approx m\,Hz$ band) from space was also briefly 
discussed in his talk. Contrarily to the earth based detectors, 
for this type of detection there is a natural advantage which is 
the fact that essentially all the noises due to the ground
are eliminated (for more detail see~\cite{Barish2}). 

For some decades since the discovery of the cosmic expansion 
by Hubble~\cite{Hublle} it was taken for granted that the
matter content of the universe was composed by the forms
of energy we can readily detect, i.e., the ordinary
matter and radiation. Such universe would expand with 
a negative expansion rate due to gravitational self-attraction 
of the matter. However, some years later, theoretical aspects 
of the stability of galaxies and some observations (motion of 
cluster of galaxies and stars and surrounding galaxies) 
indicated that most of the matter of the universe does not
emit or absorb light (dark matter). The dark matter 
clear indication found resonance in the 1980's by the inflationary
universe scenario in which the universe is taken to be flat --- the 
total energy density equal to the critical density
$\rho_c = 3\,H_0^2 / 8\pi\,G \approx 1.7 \,\, 10^{-29} g/cm^3$.
However, measurements made by that time offered a clear 
indication that the ordinary matter and the radiation 
could only account for about $10\%$ of this value --- 
inflation seemed to require dark matter.
The new consensus model became that in  which the universe 
contains primarily cold, nonbaryonic dark matter, and
although observations at that time indicated that the total 
mass density was smaller than the critical density, many 
cosmologists assumed that the flat model
was reasonably well-established. However,
further observations strongly indicated that the total matter 
density was indeed less than half of the critical density. The 
only way of having such a low mass density and a flat universe 
is if an additional, nonluminous and nonclustering ``dark energy" 
component dominates the universe today, although it must have 
been negligible in the past so as to permit structure formation. 
The general relativity Friedmann-Lema\^{\i}tre-Robertson-Walker 
(FLRW) standard modeling approach requires that this dark energy 
has a negative pressure, and the recent measurements of distant 
exploding stars (supernovae) support the existence of a 
negative-pressure dark energy such that the (total) energy density 
$\rho$ and the (total) pressure $p$ obey the relation 
$\rho+3 p<0$, which leads to an accelerating expansion for the 
universe today (for a fair list of basic references on the topics I 
have briefly mentioned here, see the very good review by Bahcall 
{\em et al.\/}~\cite{Bahcaletal}).

Luca Amendola (Osservatorio Astronomico di Roma) presented
a quite good plenary talk in which he reviewed how the 
recent independent cosmological observations constrain
the values of the cosmological density parameters, and
have indicated that the universe is filled with dark matter 
and dark energy and undergoes an accelerating expansion. 
He pointed out how recent supernovae Ia observations%
~\cite{Perlmutter,Ries} call for an acceleration expansion 
of the universe, which together with recent cosmic microwave, 
large-scale structure and lensing data give a rather strong 
indication that the matter (energy) content of the universe 
is basically composed  of dark matter ($\sim 30 \%$), 
dark energy ($\sim 70 \%$), baryonic matter (very few percent). 
For more details and references see his review paper in this 
issue~\cite{Amendola}.

Odylio D. Aguiar (INPE) told us about the status of the 
Mario Schenberg gravitation wave detector, which is
being constructed in the Institute of Physics of the
University of S\~ao Paulo, and is expected to start to 
operate in 2002. This detector has a spherical antenna,
will operate at temperature of less than $0.1 K$, and will be 
sensitive to signals of amplitude $h > 10^{-21} Hz^{-1/2}$
in the bandwidth $3.0$~--~$3.4 k Hz\,$ (for more details
on this talk see Aguiar's  paper in this issue%
~\cite{Odylio1,Odylio2}).  

It is well known that general relativity is a local metrical
theory and therefore its field equations do not fix the 
topology of the spacetime,  since geometry does not in general
dictate the topology. This freedom has resulted in a great 
deal of recent works in which the possibility that the 
universe may possess a spatial section with a non-trivial 
topology is examined. Whether we live in a finite or infinite 
space and what is the size and the shape of the universe 
are open problems of topological nature.
In his talk Helio V. Fagundes presented a brief introduction
to cosmic topology in the context of FLRW models, and
reviewed the work done by his group and collaborators 
at IFT/UNESP. The works he described range from early 
attempts to solve a controversy about quasars through 
the multiple images produced by nontrivial topologies, 
to more recent issues such as cosmic crystallography 
and on quantum cosmology in compact universe models
(for more details on this talk see Fagundes' paper 
in this issue~\cite{Fagundes}).

Before reviewing the oral presentations I remark that 
one particular objective of the organizers was to involve 
as far as possible the local general public with the meeting. 
To this end they invited I. Waga to give a public general 
lecture on cosmology, and he presented a clear and motivating 
lecture on the expansion of the universe~\cite{WagaDiv}.

\section{Oral Presentations I}

The first oral presentation was of an observationally-oriented
paper by M.O. Calv\~ao, J.R.T. de Mello Neto and I. Waga%
~\cite{CalMelWa}, where it was discussed, through Monte Carlo 
simulations, how the Alcock-Paczy\'nski test, as applied to quasar
clustering, can be used to probe the cosmological density 
parameters space, and also the equation of state parameter.
The approach discussed in this contribution may become a 
relevant strategy to narrow down the cosmological parameter 
space.

H.P. de Oliveira and S.E. Jor\'as presented an early universe 
contribution, where they studied the evolution of the entropy
and adiabatic perturbations and showed that for a small amount
of dissipation the entropy perturbations can be neglected, 
and so the purely adiabatic perturbations will be responsible
for the primordial spectrum of inhomogeneities~\cite{OliJor}. 
  
In a work presented by Sandra Roveda (M.E. Ara\'ujo, S.R.M.M. 
Roveda and  W. Stoeger~\cite{AraRovSto}) spherically symmetric
perturbations of FLRW models in the  so-called observational 
coordinates were examined. The FLRW (spherical) models are 
used as the background spacetime and a particular case of 
spherical deviation from this background is studied (for 
more details on this ``observational program" I refer 
the reader to the paper by Ellis {\em et al.}%
~\cite{Ellisetal}).

Using recent observational constraints on cosmological 
density parameters, together with recent mathematical results 
concerning small volume hyperbolic manifolds and a 
topological detectability indicator introduced in%
~\cite{grt2001a}, G.I. Gomero reported on a contribution
on cosmic topology in which it is argued that, by employing 
pattern repetitions, the topology of nearly flat small 
hyperbolic universes can be observationally 
undetectable~\cite{grt2001b}. This is an important result 
in view of the fact that quantum cosmology seems to favour 
universes with small volumes, and from the expectation, 
coming from inflationary scenarios, that the total density 
$\Omega_0$ is likely to be very close to one.

By considering that the current expanding era is preceded by
a contracting phase, in other words by assuming the existence 
of one bounce, N. Pinto-Neto showed how one can conclude 
that no observable bounce could possibly have taken place 
in the early universe if GR together with hydrodynamical 
fluids describe the evolution of the universe, thus under 
these conditions the universe has always expanded 
(see~\cite{PatrickPinto} for more details).

The first set of oral presentation was closed with the
presentation by S.E. Jor\'as of a article by R.H. Brandenberger, 
S.E. Jor\'as and J. Martin, in which the spectrum of scalar 
field fluctuations in a bouncing asymptotically flat universe 
was calculated, and the dependence of this result on length 
scales shorter than the Planck length was investigated (for more
details see~\cite{BranJorMar}).

\section{Oral Presentations II}

Topological defects such as monopoles, strings and domain
walls have been studied in different contexts such as to
understand the primordial universe and structure 
formation in the early universe. Their nature depends upon
the topology of the vacuum manifold of the field theory
under consideration. The second set of oral presentations 
began with a contribution by R.M. Teixeira Filho and 
V.B. Barbosa~\cite{TeiBar}, where they have obtained 
in the context of scalar tensor theories the gravitational 
field of a global monopole, in the weak-field approximation, 
extending the Barriola and Vilenkin monopole solution 
found in the context of GR.

S.S. e Costa delivered a contribution related to cosmic topology
in which general solutions of the Helmholtz equation, in several 
coordinates systems for two and three-dimensional hyperbolic, 
spaces were presented~\cite{SandroCosta}.

Another contribution related to cosmic topology was an article 
by D. M\"uller, H.V. Fagundes and R. Opher~\cite{MulOphFag},
where they have studied, through numerical calculations, the 
Casimir (vacuum) energy for a conformally coupled, massive 
scalar field in a static universe whose spatial sections are 
endowed with the topology of the smallest known hyperbolic 
three-manifold (Weeks manifold). They have concluded that 
there is a spontaneous vacuum excitation of low multipolar 
components.

A. Bernui also presented a cosmic topology contribution in
which the cosmic crystallography statistical approach to
the topological signature of the universe was used in his
numerical simulations with incomplete catalogs.

R. Androvandi presented an interesting contribution
(R. Aldrovandi, J. Gariel and G. Marcilhacy) on what they
call the pre-nucleosynthesis period (PNS period), between 
$z \approx 10^{10}$ and $z \approx 10^{15}$,
and where our local physics does not necessarily hold.  
A general overview of physical problems appearing in the 
PNS period was given, and possible meaning for the 
striking outcomes were discussed. For more details see%
~\cite{Aldrovandi}.

\section{Oral Presentations III}

Condensed matter systems such as light in moving dielectrics
and quasiparticles in a moving superfluid can be used to 
mimic {\em kinematic\/} aspects of general relativity. 
The comparative study between the kinematical aspects of
GR and other kinds of interactions has been called {\em
analog model\/} for gravitation or simply {\em analog
gravity\/}. These types of analog models are rather 
important in that, e.g., they provide black hole analogs 
and may lead to experimental test of quantum field theory 
in curved space.
The last set of oral presentations began with a brief
review talk, delivered by R. Klippert (a work in 
collaboration with V.A. de Lorenci) where it was discussed
analog gravity models in several contexts, ranging from 
electrodynamic in non-linear media~\cite{LorKlip} to
moving dielectrics and acoustic perturbations (for a fair
number of references on this topic see~\cite{WorkAnal}).

A work by J.M.F Maia and J.A.S. Lima in which a procedure
to generate cosmological solutions in the context of GR 
whose matter content is given by a scalar field plus a
perfect fluid was presented. Applications to inflationary 
and quintessence cosmologies were also considered in 
specific models (see~\cite{MaiaLima02}). 

A.Y. Miguelote reported on the results of an article 
in collaboration with  M.F.A. Silva, A.A. Wang and 
N.O. Santos, in which some properties of the Levi-Civita 
(LC) type of solutions of coupled Einstein-Maxwell 
equations are studied, and  some limits of this type of 
solution are obtained together with an interpretation of 
the free parameters involved in LC type of spacetimes%
~\cite{MiSiWaSa}.

I.D. Soares presented a interesting paper (homoclinic chaos 
in the dynamics of a general Bianchi IX) of a work in
collaboration with H.P. Oliveira, A.M.O. Almeida, 
and E.V. Tonini, in which they have studied the 
dynamics of anisotropic (three scale factors) 
Bianchi type IX model with dust and cosmological 
constant (positive). They have examined the existence 
of chaos in this model, and have shown that it is chaotic, 
and that the chaos has a homoclinic nature. The
role played by the cosmological constant in the phase
space is shown to be crucial in that it determines both 
the existence of a saddle-center-center critical point, and
the critical points at infinity corresponding to 
the De-Sitter configuration.  For more 
details I refer the reader to ref.~\cite{OlAlSoTo}.

The interaction of Hawking radiation and a static 
electric charge was a contribution delivered by 
L.C.B. Crispino of a work in collaboration with
A. Higuchi and G.E.A. Matsas, in which they have
investigated in two interacting situations whether 
the equality found for the response  of a static 
scalar field is maintained in the case of electric 
charges (see~\cite{LuHiMa}).

The last oral presentation  was presented by J. 
Casti\~neiras (a work in collaboration with 
L.C.B Crispino, G.E.A. Matsas and D.A.T. 
Vanzella). They have studied
``free particles" for which $E < m c^2$ outside
 Reissner-Nordstrom blackholes and in a spacetime
of a star (for more details see~\cite{LastOral}).

\section{Panel Contributions}

More than fifty contributions were presented as posters.
About twenty per cent were related to the physics of 
the early universe. F.G. Alvarenga, J.C. Fabris, G.A.
Monerat and N.A. Lemos presented two contributions on
quantum cosmology~\cite{FJNM1}. N. Pinto-Neto and
E.S. Santini also presented a poster related to quantum
cosmology (see in this connection the article~\cite{BNS}).
Three works connected to the preheating process were 
presented as panels: one by S.E. Jor\'as and V.H. 
C\'ardenas~\cite{JoCar}; a second 
by A.H. Campos, R. Rosenfeld and J.M.F. Maia; 
and another by R.O. Ramos and S.E. Jor\'as. 
Three posters were related to inflation one by H.P.
de Oliveira and I. Dami\~ao Soares, a second  by
H.P. de Oliveira alone~\cite{HPO}, and a third
by J.C. Fabris, A.M. Pelinson and I.L. Shapiro~\cite{FPS}.
A.H. Campos, R. Rosenfeld, J.M.F. Maia and H. Reis 
discussed in a poster the possible production of 
superheavy primordial particles.
R. Aldrovandi, A.L. Barbosa, M. Can\c{c}ada and
J.G. Pereira presented a poster on kinematics of a 
spacetime with infinite cosmological constant 
($\Lambda \to \infty$) and speed of light ($c \to \infty$), 
whose results may be of some interest to early universe 
cosmology~\cite{AldBarCalPer}.
A poster with some contact with the early universe
was also brought to the meeting by M.G.C. Batista.

Nine contributions related to topological defects in different
contexts were presented as posters. Three of them treated
these defects in scalar-tensor theories: one by V.B. 
Bezerra, L.P. Colatto, M.E.X. Guimar\~aes and R.M. Teixeira
Filho~\cite{BeCoGuTe}; another by V.B. Bezerra 
and C.N. Ferreira~\cite{BeFe}; and a third by V.B. Bezerra,
R.M. Teixeira Filho, G. Grebot and M.E.X. Guimar\~aes%
~\cite{BeTeGreGui}. The behaviour of nonrelativistic 
quantum particles interacting with different potentials 
in the spacetimes generated by a cosmic string and also by a 
global monopole were discussed in a panel by  G. de A. Marques
and V.B. Bezerra~\cite{GeusaValdir}.
Two posters on cosmic strings were presented: one by A.A. 
Mor\'egula in collaboration with M.M. Som, and another
by V.A. de Lorenci and E.S. Moreira Jr (see~\cite{LM}).
The solutions of Schr\"odinger equations for the Coulomb 
and harmonic oscillator potential in the cosmic-string
conical spacetimes of $N$ dimensions were presented in 
a poster by J.L.A. Coelho and R.L.P.G. Amaral~\cite{CoAma}.
Topological defects in the context of (or related to) 
condensed matter were discussed in two posters: one by 
C.A.L. Ribeiro, C. Furtado and F. Moraes, and another by 
J.F de Assis, V.B. Bezerra and C. Furtado. 

Besides Fagundes' parallel talk and the oral presentations
by G.I. Gomero, S.S. e Costa, D. M\"uller {\em et al.}, and
A. Bernui, we had an additional panel presentation by
G.I. Gomero of a work in collaboration with M.J. Rebou\c{c}as
and R. Tavakol~\cite{grt2001a} on the detectability of cosmic
topology of nearly flat FLRW universes in the light of the
most recent observations (Sne Ia, large scale structure (LSS)
observations, MAXIMA and BOOMERANG).

In addition to Barry C. Barish's plenary talk, and O.D. Aguiar's
parallel talk, four posters related to gravitational waves
were presented: K.L. Ribeiro and O.D. Aguiar; S.R. Furtado
and O.D. Aguiar; and L.A. de Andrade, O.D. Aguiar and K.L.
Ribeiro; and also C. Frajuca, N.S. Magalh\~aes, O.D. Aguiar,
K.L. Ribeiro, L.A. de Andrade, W.F. Velloso Jr and J.L.
Melo (see~\cite{Ribeiroetal,Frajucaetak}).  

Three posters were presented on parallel gravity. One
by T. Vargas on Regge Calculus in teleparallel gravity,
and three others by A.A. Sousa and J.W. Maluf (for
more details see~\cite{SousaMaluf,MalufSousa}).

It seems indisputable that most of theoretical physicists
spend a large amount of their time in carrying out nonnumeric
calculation of one sort or another, most of which are
algorithmic or semi-algorithmic. Today the computer algebra (CA)
systems are standard tools of theoretical physicists for 
performing symbolic, although in most cases perhaps they 
simply do not mention this fact. We had three CA poster 
contributions in the meeting, all related to the Maple system. 
One dealt with the symmetry groups in GR (by J.A.F. Roveda and 
J.E.F. Skea), which is a relevant piece of information needed in 
the invariant (local) characterization of spacetime; and another 
was about the usage of Maple to calculate quasi-normal modes 
of radiating systems (by P.C.C. Monteiro Jr and J.E.F. Skea). 
F.D. Sasse delivered  a contribution in collaboration with 
R.F. McLenaghan and S.R. Czapor where the Maple package 
{\tt NPspinor} was used to prove the Hadamard's conjecture 
for the scalar wave equation in Petrov type III backgrounds.

Different aspects and contexts of black holes were discussed 
in a series of four posters. G. de Oliveira Neto
studied exact solutions of Einstein's equations in 
($2+1$)-dimensions which, depending on the values of certain 
parameters, may represent the formation of black 
holes~\cite{GilONeto}.
A.A. Sobreira and V.B. Bezerra examined rotating black
hole in Einstein-Fokker theory in ($2+1$)-dimensions. 
Holonomies in the ($2+1$)-dimensional geometry of black 
holes were the content of a poster by A. Carvalho,
C. Furtado and F. Moraes. A black hole type solution
of Einstein equations in $2$-dimensions was discussed
by D.G. Delfrate.

There were four posters which I broadly grouped as 
gravitational-related works. M. Cal\c{c}ada presented
a panel on gravitation and the local symmetry group
of spacetime~\cite{calcada}. A.A. Sousa and J.W. Maluf
presented a poster where they proposed an experimental
test for the gravito-magnetic effect. R.C. Rigitano
discussed in an interesting poster the geometric 
representation of general linear transformations, and 
the possibilities and limitations of the representation 
of covariant transformations. C.R. Muniz examined 
geometric properties of the spin in the context of GR.

Two posters were related to Astrophysics. In one the 
third post-Newtonian dynamics of compact binaries
equations of motion was discussed by V. Andrade, L. 
Blanchet and G. Faye (for detail see~\cite{Vanessaetal}).
In another by K. Sau Fa and I.T. Pedron it is used Tsallis'
distribution functions to examine the galaxies in
equilibrium configurations with spherical symmetry
(King models)~\cite{FaPedron}.

M.D. Maia presented an interesting panel (related
to two works in collaboration with E.M. Monte
~\cite{MaiaEdmundo1,MaiaEdmundo2}) in which 
he showed how brane-world cosmology offers a 
quite simple explanation for dark energy.

M.E.X. Guimar\~aes, L.P. Colatto and F.B. Tourinho
presented a poster in which they extended to the
context of scalar-tensor theories~\cite{GuiColTou} 
(where the parameter $\omega$ in no longer constant) 
a procedure, devised by A. Barros and C. Romero~\cite{BR}, 
which makes clear how to obtain solutions in the Brans-Dicke
theory from solutions in GR for the same matter content 
when one considers the weak-field approximations in both 
theories.

In an interesting panel by S.O. Mendes and R. Opher it was 
discussed how the $1/r^2$ gravitational law on submillimetric 
scale can be used to test the modified Newtonian dynamic 
theory (MOND)~\cite{MendesOpher}.

Besides the brief review talk, presented by R. Klippert
as an oral presentation, Analog Gravity also appeared in a
poster by V.A. De Lorenci and R. Klippert~\cite{LorKlip}.

An interesting panel was presented by F.P. Devecchi and 
G. Kremer in which the kinetic theory of relativistic gases 
in two-dimensional space was developed to obtain 
thermodynamical quantities in $2$-dimensional cosmological
models~\cite{KramerDevecchi}. 

H.P. de Oliveira and I. Dami\~ao Soares presented in a   
contribution the results of their studies of the dynamics of
spherically symmetric gravitational collapse of a massless
scalar field~\cite{OliveiraSoares}. H. P. de Oliveira,
I. Dami\~ao Soares and E.V. Tonini also had a panel where
they presented an analytical method to describe the
unstable periodic orbits of the center manifold for
Hamiltonian systems.
Another poster with some relation to this was presented 
by O.C. Castellani, G.A. Monerat and J.F.V. Rocha. 

M.M. de Souza claimed in a panel that the constant anomalous
radial acceleration observed in the Pioneer 10 and 11 
spacecrafts can be explained in terms of discrete interactions
(for detail see~\cite{Manoelito1,Manoelito2}).

J.F. Villas da Rocha presented a poster in which it was
discussed solutions to Einstein's field equations in
$N$-dimensions with spherical symmetry~\cite{Rocha}. 

L. Sandoval Jr considered in a contribution the constraints
on generalized metrics  (with both symmetric and 
antisymmetric parts) described through vierbeins
imposed by bosonic strings in curved spacetimes.

M.L. Bedran presented an interesting pedagogical panel 
comparing the Doppler (special relativity) effects with 
the cosmological redshift of general relativity.

F.D. Sasse reviewed the different interpretations of 
the Mashhoon effect, and discussed a description of 
fermions in a storage ring in a Serret-Frenet tetrad 
frame, instead of the usual Hehl-Ni frame.

M.J. Lazo and S. Ragusa showed in  how to calculate the 
electromagnetic angular momentum emission of fourth order, 
extending therefore the calculations made in the Landau-Lifchitz 
book for electric dipole. They intend to use similar scheme 
to examine analogous gravitational emission.  

To close this section I remark that a considerable number
of extended versions of the works presented in the XXII BNMPF 
will be available in a CD (proceedings of the meeting), 
and can fairly soon be obtained from the Brazilian Physical 
Society (SBF). So, for more details on some of works briefly
reviewed here, I refer the readers to this 
CD ROM~\cite{SBFpage}.

\section{Closing Remarks}

The large number of abstracts submitted to the Cosmology and
Gravitation sessions of the XXII BNMPF is certainly an 
indicative of the robust level of interest in the area.
The meeting was a lively one and demonstrates the vitality
of a extensive set of research topics. It also was 
very informative and many discussions took place in the regular
sessions, and also in the coffee and others breaks. The
spirit of collaboration, essential to the progress of
science, was present in the sessions and discussions.

Let me close my review by strengthening a point I have 
indicated in the introduction, which is that the interaction 
and collaboration between theoretical and observationally-oriented 
gravito-cosmologists is strongly advisable for the advance of the
area, the benefit of our community as well as for the formation 
of Brazilian young scientists in Cosmology and Gravitation.

\begin{acknowledgments}
I would like to thank the Organizing Committee for the invitation
to deliver the summary talk on Gravitation and Cosmology of
the XXII BNMPF. I also thank them for the large amounts of
effort put into the organization of a very successful meeting. 
I thank Ioav Waga for his help in the classification of the
topics for my summary talk.
Some authors provided me with preprints, reprints or references
related to their submitted abstracts. The additional information 
material proved to be very useful for me, and I thank all these 
colleagues for their helpful gesture.
I also thank Antonio F.F. Teixeira and G.I. Gomero for reading 
the manuscript and for the useful suggestions.
Finally, I  thank CNPq for the grant under which this work 
was carried out.
\end{acknowledgments}


\begin{references}

\bibitem{IWaga1} I. Waga, {\em Braz.\ J. Phys.\/} {\bf 31},
285 (2001). 

\bibitem{Opher1} See the of abstracts of contributed papers,
{\em Workshop: New Physics in the Space\/}, held in Campos 
do Jord\~ao, March 03 to 08, 2002, and organized by R. Opher 
(IAG/USP), C. Escobar (IF/UNICAMP), G. Matsas (IFT/UNESP), 
O. Aguiar (DAS/INPE), P. Pellegrini (ON), R. Rosenfeld(IFT/UNESP),
S.M. Viegas (IAG/USP), T. Villela (DAS/INPE) and Z. Abraham 
(IAG/USP). 

\bibitem{TaylorRev} J.H. Taylor, {\em Rev.\ of Mod.\ Phys.\/}
{\bf 66}, 711 (1994). See also references therein.

\bibitem{Will} C.M. Will, {\em Theory and Experiment in Gravitational
Physics\/}. Cambridge University Press (1981).

\bibitem{Barish1} B.C. Barish, {\em Gravitational Waves the
New Generations of Laser Interferometric Detectors\/}, in
''Proceedings of the Ninth Marcel Grossmann Meeting on General 
Relativity (2000), V. Gurzadyan, R.T. Jantzen, R. Ruffini, Eds.,
World Scientific, Singapore, 2002. Available at 
http://www.icra.it/MG/mg9/mg9.htm

\bibitem{Barish2} B.C. Barish, {\em Braz.\ J. Phys.\/} , this 
issue (2002).

\bibitem{Hublle} E. Hubble, {\em Proc.\  Nat.\ Acad.\ Sci.\/}
{\bf 15}, 168 (1929).

\bibitem{Bahcaletal} N.A. Bahcall, J.P. Ostriker, S. Perlmutter
and P.J. Steinhardt, {\em Science} {\bf 284}, 1481 (1999).

\bibitem{Perlmutter} S. Perlmutter et al., {\em Astrophys.\
J.\/} {\bf 517}, 565 (1999).

\bibitem{Ries} A.G. Riess et al., {\em Astron.\
J.\/} {\bf 116}, 1009 (1998).

\bibitem{Amendola} L. Amendola, {\em Braz.\ J. Phys.\/} , this 
issue (2002).
 
\bibitem{Odylio1} O.D. Aguiar, {\em Braz.\ J. Phys.\/} , this 
issue (2002).

\bibitem{Odylio2} O.D. Aguiar {\em et al.\/},
{\em Class. Quantum Grav.\/} {\bf 19}, 1949 (2002).

\bibitem{Fagundes} H.V. Fagundes, {\em Braz.\ J. Phys.\/} , this 
issue (2002). Also preprint gr-qc/0112078 (2001).

\bibitem{WagaDiv} I. Waga, the text of this lecture is
available at http://www.if.ufrj.br/~ioav/nota.html. 

\bibitem{CalMelWa} M.O. Calv\~ao, J.R.T. Mello Neto and
I. Waga, {\em Phys.\ Rev.\ Lett.\/} {\bf 88}, 091302 (2002). 

\bibitem{OliJor} H.P. Oliveira and S.E. Jor\'as, {\em Phys.\
Rev.\ D\/} {\bf 64}, 063513 (2001).

\bibitem{AraRovSto} M.E. Ara\'ujo, S.R.M.M. Roveda and W. Stoeger,
``Perturbed Spherically Symmetric Dust Solution of the Field
Equations in Observational Coordinates with Cosmological Data
Functions", preprint gr-qc/0105001 (2001).

\bibitem{Ellisetal} G.F.R. Ellis, S.D. Nel, R. Maartens, 
W.R. Soeger and A.P. Whitman, {\em Phys.\ Rep.\/} {\bf 124}
315 (1985).

\bibitem{grt2001a} G.I. Gomero, M.J. Rebou\c{c}as and R. Tavakol,
{\em Class.\ Quantum Grav.\/} {\bf 18}, 4461 (2001).

\bibitem{grt2001b} G.I. Gomero, M.J. Rebou\c{c}as and R. Tavakol,
{\em Class.\ Quantum Grav.\/} {\bf 18}, L145 (2001).

\bibitem{PatrickPinto} P. Peter, N. Pinto-Neto, {\em Phys.\
Rev.\ D\/} {\bf 65}, 023513 (2002).

\bibitem{BranJorMar} R.H. Brandenberger, S.E. Jor\'as, J. Martin,
``Trans-Planckian Physics and the Spectrum of Fluctuations in a 
Bouncing Universe", preprint hep-th/0112122 (2001).

\bibitem{TeiBar} R.M. Teixeira Filho and V.B. Bezerra,
{\em Phys.\ Rev.\ D\/} {\bf 64} 067502 (2001).

\bibitem{SandroCosta} S.S. e Costa, ``A Description of 
Several Coordinate Systems for Hyperbolic Spaces",
preprint math-ph/0112039 (2001).

\bibitem{Aldrovandi} R. Aldrovandi, J. Gariel and G. Marcilhacy,
``On the Pre-nucleosynthesis Cosmological Period", preprint 
gr-qc/0203079 (2002).

\bibitem{MulOphFag} D. M\"uller, H.V. Fagundes and R. Opher,
{\em Phys.\ Rev.\ D\/} {\bf 63}, 123508 (2001).

\bibitem{LorKlip} V.A. De Lorenci and R. Klipert,
{\em Phys.\ Rev.\ D\/} {\bf 65}, 064027 (2002).

\bibitem{WorkAnal} See the workshop web page on analog models of
general relativity at http://www.cbpf.br/$\sim$bscg/analog

\bibitem{MaiaLima02} J.M.F Maia and J.A.S. Lima,
{\em Phys.\ Rev.\ D\/} {\bf 63}, 083513 (2002). 

\bibitem{MiSiWaSa} A.Y. Miguelote, M.F.A. da Silva, A. Wang,
and N.O. Santos, {\em Class.\ Quant.\ Grav.\/} {\bf 18}, 4569 
(2001).
 
\bibitem{OlAlSoTo} H.P. de Oliveira, A.M. Oz\'orio de Almeida, 
and I. Dami\~ao Soares, E.V. Tonini, ``Homoclinic Chaos in the Dynamics 
of a General Bianchi IX Model", preprint gr-qc/0202047(2002), to
appear in {\em Phys.\ Rev.\ D\/} (2002).

\bibitem{LuHiMa} L.C.B. Crispino, A. Higuchi and G.E.A. Matsas,
{\em Phys.\ Rev.\ D\/} {\bf 58}, 084027 (1998).

\bibitem{LastOral} J. Casti\~neiras, L.C.B. Crispino, G.E.A. 
Matsas and D.A.T. Vanzella, ``Free Massive Particles with Total
Energy $E < m c^2$ in Curved Spacetimes", preprint 
gr-qc/0201093 (2002).

\bibitem{FJNM1} F.G. Alvarenga, J.C. Fabris, N.A. Lemos and
G.A. Monerat, ``Quantum Cosmological Perfect Fluid Models",
preprint gr-qc/0106051. To appear in 
{\em Gen.\ Rel.\ Grav.\/} (2002).

\bibitem{BNS} J.A. de Barros, N. Pinto-Neto and M.A. 
Sagioro-Leal  {\em Phys.\ Lett.\  A\/} {\bf 241}, 229 (1998).

\bibitem{JoCar} S.E. Jor\'as and V.H. C\'ardena,
``Chaos and Preheating", preprint gr-qc/0108088 (2001).

\bibitem{HPO} H.P. de Oliveira, {\em Phys.\ Lett.\ B\/} 
{\bf 526}, 1 (2002).

\bibitem{FPS} J.C. Fabris, A.M. Pelinson and I.L. Shapiro,
{\em Nucl.\ Phys.\  B\/} {\bf 597}, 539 (2001); Erratum-ibid. 
{\bf 602}, 644 (2001). 

\bibitem{AldBarCalPer}  R. Aldrovandi, A.L. Barbosa, 
M. Can\c{c}ada and J.G. Pereira, ``Kinematics of a
Spacetime with Infinite $\Lambda$ and $c$, preprint
gr-qc/0105068 (2001).


\bibitem{BeCoGuTe} V.B. Bezerra, L.P. Colatto,
M.E.X. Guimar\~aes and R.M. Muniz Filho, ``Scalar and 
Spinor Particles in the Spacetime of a Domain Wall 
in String Theory", {\em Phys.\ Rev. D\/} {\bf 65},
in press (2002), also preprint gr-qc/0104038 (2001).

\bibitem{BeFe} V.B. Bezerra and C.N. Ferreira,
{\em Phys.\ Rev.\ D\/} {\bf 65}, in press (2002),
also preprint hep-th/0111167 (2001).
 
\bibitem{BeTeGreGui}  V.B. Bezerra, R.M. Teixeira Filho,
G. Grebot and M.E.X. Guimar\~aes, {\em Mod.\ Phys.\
Lett.\ A\/} {\bf 16}, 1565 (2001).

\bibitem{GeusaValdir} G. de A. Marques and V.B. Bezerra,
{\em Class.\ Quantum Grav.\/} {\bf 19}, 985 (2002).

\bibitem{LM} V.A. De Lorenci and E.S. Moreira Jr,
{\em Phys.\ Rev.\ D\/} {\bf 63}, 027502 (2001).

\bibitem{CoAma} J.L.A. Coelho and R.L.P.G. Amaral,
``Coulomb and Quantum Oscillator Problems in Conical 
Spaces with Arbitrary Dimensions", preprint 
gr-qc/0111114 (2001).

\bibitem{Ribeiroetal}  K.L. Ribeiro {\em et al.\/},
{\em Class.\ Quantum Grav.\/} {\bf 19}, 1967 (2002).

\bibitem{Frajucaetak} C. Frajuca {\em et al.\/}
{\em Class.\ Quantum Grav.\/} {\bf 19}, 1961 (2002).

\bibitem{SousaMaluf} A.A. Sousa and J.W. Maluf,
{\em Prog.\ Theor.\ Phys.\/} {\bf 104}, 531 (2000).

\bibitem{MalufSousa} J.W. Maluf and A.A. Sousa,
``Hamiltonian Formulation of Teleparallel Theories 
of Gravity in the Time Gauge", preprint 
gr-qc/0002060 (2000).

\bibitem{GilONeto} G. de Oliveira Neto, ``Self-similar
Collapse of a Massless Scalar Field in Three-dimensions",
preprint gr-qc/0105077 (2001).

\bibitem{calcada} M. Cal\c{c}ada, {\em Int.\ J.\ 
Theor.\ Phys.\/} {\bf 41}, 729 (2002).

\bibitem{Vanessaetal} V. Andrade, L. Blanchet and G. Faye,
{\em  Class.\ Quant.\ Grav.\/} {\bf 18}, 753 (2001).

\bibitem{FaPedron} K. Sau Fa and I.T. Pedron,
``Extended King Models for Star Clusters", 
preprint astro-ph/0108370.

\bibitem{MaiaEdmundo1} M.D. Maia, E.M. Monte,
``Geometirc Stability of Brane-worlds", preprint
hep-th/0103060 (2001).

\bibitem{MaiaEdmundo2} M.D. Maia, E.M. Monte,
``Geometry of Brane-Worlds", preprint
hep-th/0110088 (2001).

\bibitem{GuiColTou} M.E.X. Guimar\~aes, L.P. Colatto
and F.B. Tourinho, {\em Spacetime and Substance\/} 
{\bf 2}, 71 (2001).

\bibitem{BR} A. Barros and C. Romero, {\em Phys.\ Lett.\ 
A\/} {\bf 245}, 31 (1998).

\bibitem{MendesOpher} S.O. Mendes and R. Opher,
{\em Phys.\ Lett.\  B\/} {\bf 522}, 1 (2001).

\bibitem{KramerDevecchi} G.M. Kremer and F.P. Devecchi,
``Thermodynamics and Kinetic Theory of Relativistic Gases 
in 2-D Cosmological Models", to appear in {\em Phys.\ Rev.\ 
D\/} {\bf 65} (2002), also preprint gr-qc/0202025 (2002).


\bibitem{OliveiraSoares} H.P. de Oliveira and I. Dami\~ao Soares,
{\em Phys.\ Rev.\  D\/} {\bf 65}, 064029 (2002). 

\bibitem{Manoelito1} M.M. de Souza,``Gravity and Antigravity 
with Discrete Interactions: Anternatives I and II", preprint 
gr-qc/0111052 (2001).

\bibitem{Manoelito2} M.M. de Souza, ``Discrete Interactions 
and the Pioneer Anomalous  Acceleration: Alternative II", 
preprint gr-qc/0106047 (2001).

\bibitem{Rocha}  J.F.V. Rocha, {\em Int.\ J.\ Mod.\ 
Phys.\ D\/} {\bf 11}, 113 (2002).

\bibitem{SBFpage} See Brazilian Physical Society (SBF) web 
page at http://www.sbf.if.usp.br


\end{references}
\end{document}